\documentclass[preprint,12pt,a4paper]{article}
\usepackage{amssymb}
\usepackage{amsmath}
\usepackage{amsfonts}
\usepackage{amssymb}
\usepackage{graphicx}
\usepackage{url}

\begin{document}

\sloppy\lineskip=0pt

\title{Turing Patterns with Turing Machines:\\Emergence and Low-level Structure Formation}

\author{Hector Zenil\footnote{h.zenil@sheffield.ac.uk}\\Behavioural and Evolutionary Theory Lab\\Department of Computer Science\\The University of Sheffield, UK}

\date{}

\maketitle

\begin{abstract}
Despite having advanced a reaction-diffusion model of ODE's in his 1952 paper on morphogenesis, reflecting his interest in mathematical biology, Alan Turing has never been considered to have approached a definition of Cellular Automata. However, his treatment of morphogenesis, and in particular a difficulty he identified relating to the uneven distribution of certain forms as a result of symmetry breaking, are key to connecting his theory of universal computation with his theory of biological pattern formation. Making such a connection would not overcome the particular difficulty that Turing was concerned about, which has in any case been resolved in biology. But instead the approach developed here captures Turing's initial concern and provides a low-level solution to a more general question by way of the concept of algorithmic probability, thus bridging two of his most important contributions to science: Turing pattern formation and universal computation. I will provide experimental results of one-dimensional patterns using this approach, with no loss of generality to a $n$-dimensional pattern generalisation.\\

\end{abstract}

\section{Introduction}

Much is known today about how pattern formation is accomplished in biology, and moreover about how this relates to Turing's work. For instance, the conditions that must be satisfied for pattern formation to occur are very clear. Different diffusion rates alone are not sufficient, and it is well understood why patterns are reproducible even if initiated by random fluctuations. Today we also know that self-assembly is a process that differs from this type of pattern formation, and we understand why, as a rule, early steps in development are usually not based on symmetry breaking, although the biochemical machinery would still be able to produce patterns under such conditions. 

This paper discusses the role of algorithmic probability in building a bridge between Turing's key scientific contributions on pattern formation and universal computation. After discussing various aspects of pattern formation in biology, cellular automata, and Turing machines, an approach based on algorithmic information theory is introduced, and experimental results relating to the complexity of producing one-dimensional patterns by running Turing machines with 2 symbols are presented. Thus the paper reconnects Turing's work on morphogenesis with his work on Turing universality by way of algorithmic probability as the theory of pattern formation at the lowest level.

\subsection{Turing patterns}

Turing provided a mechanistic mathematical model to explain features of pattern formation using reaction-diffusion equations, while coming close to achieving a first definition of cellular automata. In a recent paper Wolfram \cite{wolframturing} asks whether perhaps Turing had Turing machines in mind when developing his model of morphogenesis. Coincidentally, while Turing was working on his paper on pattern formation \cite{turingmorpho} (received in 1951, published in 1952), Niels Barricelli was performing some digital simulations in an attempt to understand evolution with the aid of computer experiments \cite{barricelli,barricelli2,dyson}, and a year after the publication of Turing's paper, the team led by Watson and Crick made a groundbreaking discovery (1953), viz. the double-helical structure of DNA \cite{dna}. Had Turing known about DNA as a biological molecule serving as memory in biological systems, carrying the instructions for life, he may have grasped the remarkable similarity between DNA and his machine tapes \cite{turing}. 

For a central element in living systems happens to be digital: DNA sequences refined by evolution encode the components and the processes that guide the development of living organisms. It is this information that permits the propagation of life. It is therefore natural to turn to computer science, with its concepts designed to characterise information (and especially digital information, of particular relevance to the study of phenomena such as DNA), but also to computational physics, in order to understand the processes of life.

Central to Turing's discussion of pattern formation is the concept of symmetry breaking, which, however, is bedevilled by a certain difficulty that Turing himself underscored, viz. how to explain the uneven distribution of biological threats versus the random disturbances triggering his machinery of reaction-diffusion. This difficulty will be generalised through the use of the so-called \emph{coding theorem} \cite{calude,cover} relating the frequency (or multiplicity) of a pattern to its (algorithmic) complexity, and what is known as Levin's Universal Distribution, at the core of which is the concept of computational universality and therefore the Turing machine. Thus we will be connecting two of Turing's most important contributions to science. I will propose that a notion of emergence can be captured by algorithmic information theory, matching identified features of emergence such as irreducibility with the robustness of persistent structures in biology. This amounts to suggesting that part of what happens, even in the living world, can be understood in terms of Turing's most important legacy: the concept of universal computation. 

Formally, a Turing machine can be described as follows: $M = (Q \cup {H}, \Sigma, \delta)$, where $Q$ is the finite set of (non-halting) states and $H$ an identified (halting) state, $\Sigma$ is the finite set of symbols (including the blank symbol 0), and $\delta$ is the next move function defined as: $\delta : Q \times \Sigma \rightarrow (\Sigma \times Q \cup H \times \{L, R\})$. If $\delta(s, q) = (s^\prime, q^\prime, D)$, when the machine is in state $q\in Q$ and the head reads symbol $s\in \Sigma$, the  machine $M$ replaces it with $s^\prime \in \Sigma$, changes to state $q^\prime \in \Sigma \cup H$, and moves in direction $D \in \{L, R\}$ ($L$ for left and $R$ for right).

\subsection{Uneven distribution of biological forms}

In the pursuit of his interest in biological pattern formation, Turing identified symmetry breaking as key to the process behind the generation of structure. The early development of, for example, an amphibian such as a frog is initiated by fertilisation of an egg and a sequence of cell divisions that result in something called a blastula. At some point the blastula acquires an axis of symmetry and one can speak of the organism's poles. So in the early stages of development, the blastula cells cease to be identical and acquire differing characteristics, ultimately constituting different parts in the developed organism. This process of differentiation of a group of cells became the focus of Turing's interest. However, biological forms, as Turing notes, are not uniformly distributed, a difficulty that he believed required an explanation \cite{turingmorpho}:

\begin{quote}
There appears superficially to be a difficulty confronting this theory of morphogenesis, or, indeed, almost any other theory of it. An embryo in its spherical blastula stage has spherical symmetry, or if there are any deviations from perfect symmetry, they cannot be regarded as of any particular importance, for the deviations vary greatly from embryo to embryo within a species, though the organisms developed from them are barely distinguishable. One may take it therefore that there is perfect spherical symmetry. But a system which has spherical symmetry, and whose state is changing because of chemical reactions and diffusion, will remain spherically symmetrical forever (The same would hold true if the state were changing according to the laws of electricity and magnetism, or of  quantum mechanics.). It certainly cannot result in an organism such as a horse, which is not spherically symmetrical. There is a fallacy in this argument. It was assumed that the deviations from spherical symmetry in the blastula could be ignored because it makes no particular difference what form of asymmetry there is. It is, however, important that there are \emph{some} [sic] deviations, for the system may reach a state of instability in which these regularities, or certain components of them, tend to grow. If this happens a new and stable equilibrium is usually reached, with the symmetry entirely gone.
\end{quote}

The phenomenon of symmetry breaking is central in Turing's discussion, as it is apparently the only possible explanation of the cause of the instability needed to start off his mechanism of pattern formation from random disturbances. But it also presents a certain difficulty that Turing himself identified as the problem of the uneven distribution of biological properties from random disruptions (in his particular example, bilateral symmetry). Turing reasoned that in the earliest stages of cell division, essentially identical sub-units were being created. But eventually this homogeneous state gave way to patterns, resulting from differentiation. In the next section (Left-handed and right-handed organisms) of Turing's paper \cite{turingmorpho}, he identifies the paradox of using random disturbances as the ignition for pattern formation, taking as an example the morphological asymmetries of organisms and species:

\begin{quote}
The fact that there exist organisms which do not have left-right symmetry does not in itself cause any difficulty. It has already been explained how various kinds of symmetry can be lost in the development of the embryo, due to the particular disturbances (or `noise') influencing the particular specimen not having that kind of symmetry, taken in conjunction with appropriate kinds of instability. The difficulty lies in the fact that there are species in which the proportions of left-handed and right-handed types are very unequal.
\end{quote}

Turing himself provides some clues as to why such a bias towards certain asymmetries would arise. It is reasonable to expect that all manner of constraints shape the way in which symmetries occur, from physical to chemical and of course biological forces (natural selection clearly being one of them). But one needs to take into account that some of these disturbances may be attenuated or amplified by physical, chemical or biological constraints, producing asymmetries. Gravity, for example, is a non-symmetric force (it always pulls from the outside toward the centre of the earth) that imposes a clear constraint (animal locomotion is therefore always found to be towards the surface of the earth). Today we know that parity violation is common, and not only in the biological world---the surrounding physical world has a bias towards matter (as opposed to anti-matter), and it is well known that explaining how an homochirality imbalance towards left-handed molecules arises is difficult (see \cite{meierhenrich}). But just as physical, chemical and biological forces impose constraints on the shapes organisms may assume, the informational character of the way in which organisms unfold from digital instructions encoded in DNA means that they must also be subject to informational and computational principles, one of which determines the frequency distribution of patterns in connection to their algorithmic complexity.

\begin{figure}[h!]
\centering
\scalebox{.505}{\includegraphics{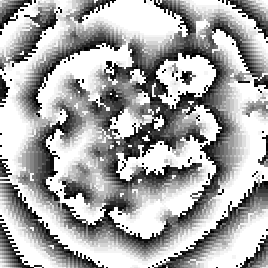}}
  \hspace{30pt}
\scalebox{.36}{\includegraphics{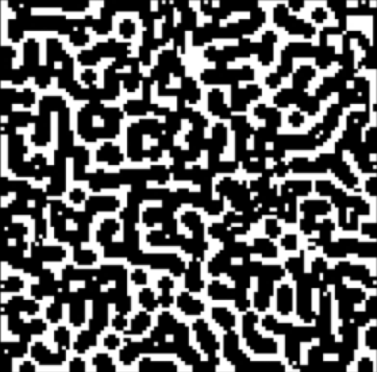}}
\caption{\small Turing patterns exhibited in simple 2D binary-state totalistic and semi-totalistic CA. In this case patterns generated by so-called \emph{Generations} CA with inhibition-activation Turing type oscillatory properties. Plotted here are the so-called \emph{Bloomerang} (left) and \emph{R(1616)} (right), found by John Elliott, the latter studied in \cite{genaro}.}
\end{figure}

The amphibian embryo mentioned above represents a signal instance where symmetry breaking is circumvented: one axis is already irreversibly determined before fertilisation takes place, and a second axis is fixed by the entry of the sperm, which initiates an intricate chain of downstream events. We stress that the aim of this paper is not to confront these problems in biology. A biologist interested in Turing's mechanism may think it would hardly profit from more formal connections or a better understanding of how a Turing machine works. However, I think that algorithmic probability, which is based on Turing's notion of universality, does have great relevance to biology, being potentially capable of explaining aspects of designability and robustness \cite{kamalzenil}.

 That Turing himself wasn't able to make a direct connection between his ideas on biology and computation at a deeper level is understandable. Here I will make a connection by way of the theory of algorithmic information, a theory that was developed a decade after Turing's paper on morphogenesis. The proposed generalisation reconnects Turing's machines to Turing's patterns.

\section{Turing's computational universality}
 
The basic finding is that there is a cyclical dynamic process that some chemicals are capable of, so that they inhibit and reactivate each other in a quasi-periodic fashion. The essential value of Turing's contribution lies in his discovery that simple chemical reactions could lead to stable pattern formation by first breaking the symmetry of the stable chemical layer with another chemical. Morphogens, as substances capable of interacting and producing patterns, had been studied since the end of the 19th. century \cite{lawrence}, but what matters in Turing's model isn't the particular identity of the chemicals, but how they interact in a mechanical fashion modelled by a pair of equations, with concentrations oscillating between high and low and spreading across an area or volume. 

The specific equations that Turing advanced do not apply to every pattern formation mechanism, and researchers have derived other formation processes either by extending Turing's framework or in various other ways (see \cite{maini}). 

I wish, however, to bridge Turing's work and his other seminal contribution to science: universality. This by way of a low-level explanation of pattern formation, a purely computational approach to pattern formation that was not available in Turing's day. The theory encompasses all kinds of patterns, even outside of biology.


Turing's most important contribution to science was his definition of universal computation, an attempt to mechanise the concept of a calculating machine (or a computer). A universal (Turing) machine is an abstract device capable of carrying out any computation for which an instruction can be written. More formally, we say that a Turing machine $U$ is universal if for an input $s$ for a Turing machine $M$, $U$ applied to $(<M>,s)$ halts if $M$ halts on $s$ and provides the same result as $M$, and does not halt if $M$ does not halt for $s$. In other words, $U$ simulates $M$ for input $s$, with $M$ and $s$ an arbitrary Turing machine and an arbitrary input for $M$.

 The concept formed the basis of the digital computer and, as suggested in \cite{brenner}, there is no better place in nature where a process similar to the way Turing machines work can be found than in the unfolding of DNA transcription. For DNA is a set of instructions contained in every living organism empowering the organism to self-replicate. In fact it is today common, even in textbooks, to consider DNA as the digital repository of the organism's development plan, and the organism's development itself is not infrequently thought of as a mechanical, computational process in biology.

\subsection{Turing's anticipation of a definition of Cellular Automata}

The basic question Turing raised in his morphogenesis paper \cite{turingmorpho} concerned the way in which cells communicated with each other in order to form structures. In proposing his model he laid down a schema that may seem similar to cellular automata in several respects. 

The theory of cellular automata is a theory of machines consisting of cells that update synchronously at discrete time steps. The earliest known examples were two-dimensional and were engineered with specific ends in view (mainly having to do with natural phenomena, both physical and biological) and are attributed to Ulam \cite{ulam} and von Neumann \cite{neumann}. The latter considered a number of alternative approaches between 1948 and 1953 before deciding on a formulation involving cellular automata. 2D cellular automata are even more relevant to natural systems because biology is essentially a two-dimensional science. One almost always thinks about and studies biological objects such as cells and organs as two-dimensional objects. Life is cellular (in the biological sense), and cells build organs by accumulating in two-dimensional layers (we are basically a tube surrounding other tubular organs, and the development of all these organs is also basically two-dimensional). Among cellular automata there is one particular kind studied by Wolfram \cite{wolfram} called Elementary Cellular Automata (ECA), because by most, if not any standards, they constitute the simplest rulespace set. An ECA is a one-dimensional CA the cells of which can be in two states, and which update their states according to their own state and the states of their two closest neighbours on either side. 

Formally, an elementary cellular automaton (ECA) is defined by a local function $f:\{0, 1\}^3 \rightarrow \{0, 1\}$, 
which maps the state of a cell and its two immediate neighbours to a new cell state. There are $2^{2^3} =
256$ ECAs and each of them is identified by its Wolfram number $\omega = \sum_{a,b,c \in {0,1}} 2^{4a+2b+c}f(a,b,c)$ \cite{wolfram}.

 It is in fact clear that the once commonly held belief about automata, viz. that complex behaviour required complex systems (with a large number of components) derived from von Neuman's \cite{neumann}. But this was soon falsified by the work of, for example, Minsky \cite{minsky}, and generally disproved by Wolfram \cite{wolfram} with his systematic minimalistic approach.

Turing's problem, however, was completely different, as it was not about self-replication but about producing patterns from simple components, except that Turing would describe the transition among cells by way of ordinary differential equations. But one of Turing's two approaches was very close to the current model of CA, at least with respect to some of its most basic properties (and not far from modern variations of \emph{continuous} Cellular Automata).


However, it was von Neumann who bridged the concepts of Turing universality and self-replication using his concept of the universal constructor, giving rise to the model of CA, and demonstrated that the process was independent of the constructor. This was not trivial, because the common belief was that the constructor had to be more complicated than the constructed. von Neumann showed that a computer program could contain both the constructor and the description of itself to reproduce another system of exactly the same type, in the same way that Turing found that there were Turing machines that were capable of reproducing the behaviour of any other Turing machine. 

Among the main properties of cellular automata (CA) as a model of computation (as opposed to, for example, Turing machines) is that their memory relies on the way states depend on each other, and on the synchronous updating of all cells in a line in a single unit of time. 

For a one-dimensional CA (or ECA) the evolution of the cells change their states synchronously according to a function $f$. After time $t$ the value of a cell depends on its own initial state together with the initial states of the $N$ immediate left and $n$ immediate right neighbour cells. In fact, for $t = 1$ we define $f^1(r_{-1},r_0,r_1) = f(r_{-1},r_0,r_1)$. 

An immediate problem with defining the state of a cell on the basis of its own state and its neighbours' states is that there may be cells on the boundaries of a line which have no neighbours to the right or the left. This could be solved either by defining special rules for these special cases, or as is the common practice, by configuring these cells in a circular arrangement (toroidal for a 2D ECA), so that the leftmost cell has as its neighbour the rightmost cell. This was the choice made by Turing. To simplify his mathematical model, he only considered the state's chemical component, such as the chemical composition of each separate cell and the diffusibility of each substance between two adjacent cells. Since he found it convenient to arrange them in a circle, cell $i=N$ and $i=0$ are the same, as are $i = N + 1$ and $i = 1$. In Turing's own words \cite{turingmorpho}:

\begin{quote}
One can say that for each $r$ satisfying $1\leq r \leq N$ cell $r$ exchanges material by diffusion with cells $r-1$ and $r+1$. The cell-to-cell diffusion constant from $X$ will be called $\mu$, and that for $Y$ will be called $\nu$. This means that for a unit concentration difference of $X$, this morphogen passes at the rate $\mu$ from the cell with the higher concentration to the (neighbouring) cell with the lower concentration.
\end{quote} 

Turing's model was a circular configuration of similar cells (forming, for example, a tissue), with no distinction made among the cells, all of which were in the same initial state, while their new states were defined by concentrations of biochemicals. The only point of difference from the traditional definition of a cellular automaton, as one can see, is in the transition to new states based on the diffusion \emph{function} $f=\mu,\nu$ which dictates how the cells interact, satisfying a pair of ordinary differential equations (ODEs). Just as in a traditional cellular automaton, in Turing's model the manner in which cells would interact to update their chemical state involved the cell itself and the closest cell to the right and to the left, resulting in the transmission of information in time.

 Turing studied the system for various concentrations satisfying his ODEs. He solved the equations for small perturbations of the uniform equilibrium solution (and found that his approach, when applied to cells and points, led to indistinguishable results). Turing showed that there were solutions to his ODEs governing the updating of the cells for which stable states would go from an initial homogeneous configuration at rest to a diffusive instability forming some characteristic patterns. The mathematical description is given by several ODEs but Turing was clearly discretising the system in units, perhaps simply because he was dealing with biological cells. But the fact that the units were updated depending on the state of the cell and the state of neighbouring cells brings this aspect of Turing's work close to the modern description of a cellular automaton.



That the mechanism could produce the same patterns whether the units were cells or points suggests that the substratum is less important than the transition mechanism and, as Turing pointed out, this wouldn't be surprising if the latter situation is thought of as a limiting case of the former. The introduction of cells, however, may have also been a consequence of his ideas on computation, or else due to the influence of, for example, von Neumann.

 \subsection{Inhibition and activation oscillatory phenomena}


 Turing was the first to realise that the interaction of two substances with different diffusion rates could cause pattern formation. An excellent survey of biological pattern formation is available in \cite{koch}. At the centre of Turing's model of pattern formation is the so-called reaction-diffusion system. It consists of an ``activator,'' a chemical that can produce more of itself; an ``inhibitor'' that slows production of the activator; and a mechanism diffusing the chemicals. A question of relevance here is whether or not this kind of pattern formation is essentially different from non-chemically based diffusion. It turns out that Turing pattern formation can be simulated with simple rule systems such as cellular automata. 
In 1984, Young \cite{young} proposed to use Cellular Automata to simulate the kind of reaction-diffusion systems delineated by Turing. He considered cells laid out on a grid in two states (representing pigmented and not pigmented). The pigmented cell was assumed to produce a specified amount of activator and a specified amount of inhibitor that diffused at different rates across the lattice. The status of each cell changed over time depending on the rules, which took into account the cell's own behaviour and that of its neighbours. Young's results were similar to those obtained using continuous reaction-diffusion equations, showing that this behaviour is not peculiar to the use of ODEs.

\begin{figure}[h!]
\centering
\scalebox{.32}{\includegraphics{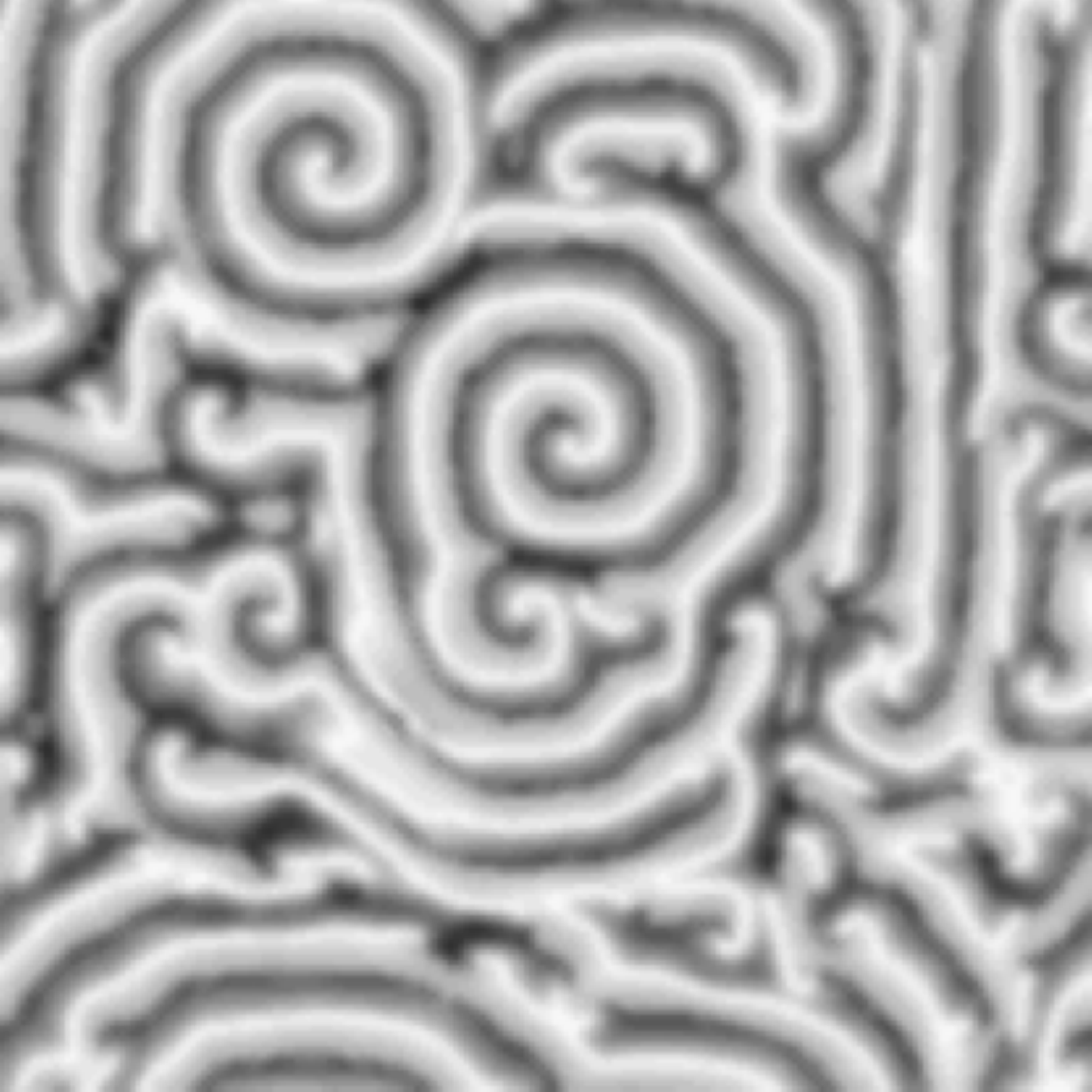}}
  \hspace{30pt}
\scalebox{.4}{\includegraphics{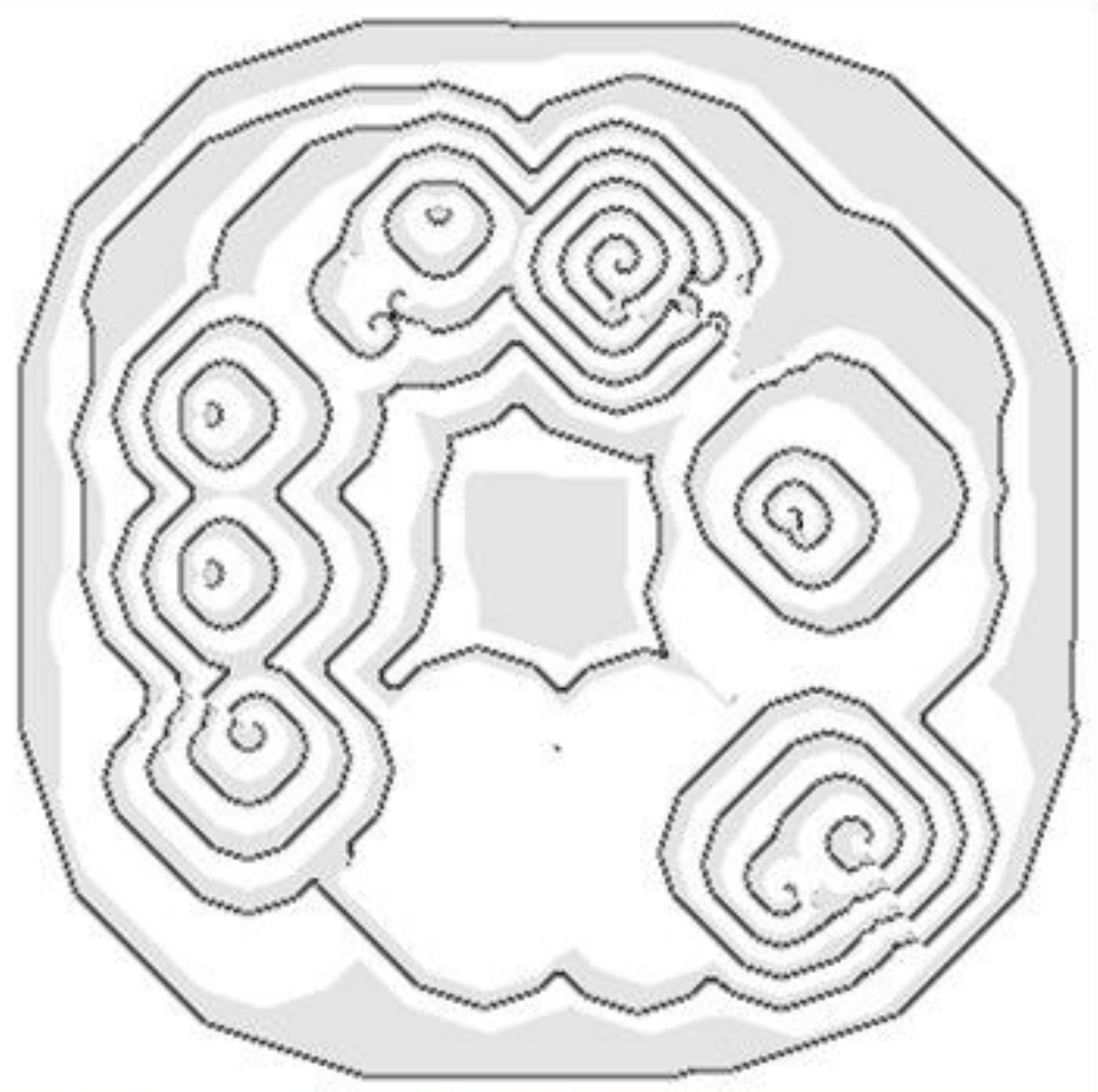}}
\caption{\small Mathematical simulation of a Belouzov-Zhabotinsky reaction (left) \cite{belousov}. The Belousov-Zhabotinsky chemical reaction pattern can be implemented by means of a 2D totalistic CA (right) with a range 1, 20-colour totalistic CA \cite{zammataro}. In the same rule space of 2D CA a wide number of rules produce Belousov-Zhabotinsky patterns.}
\end{figure}

Seashells, for example, provide a unique opportunity because they grow one line at a time, just like a one-dimensional (1D) cellular automaton, just like a row of cells in Turing's model. The pigment cells of a seashell reside in a band along the shell's lip, and grow and form patterns by activating and inhibiting the activity of neighbouring pigment cells. This is not an isolated example. Seashells seem to mimic all manner of cellular automata patterns \cite{meinhardt,wolfram}, and moving wave patterns can be simulated by 2D cellular automata, producing a wide range of possible animal skin patterns.



 In \cite{genaro}, reaction/diffusion-like patterns in CA were investigated in terms of space-time dynamics, resulting in the establishment of a morphology-based classification of rules. The simplest possible model of a quasi-chemical system was considered, based on a 2D CA and Moore neighbourhood with two states or colours. As defined in \cite{genaro}, every cell updates its state by a piecewise rule.

 In this type of cellular automaton a cell in state 0 assumes state 1 if the number of its neighbours in state 1 falls within a certain interval. Once in state 1 a cell remains in this state forever. In this way the model provides a substrate and a reagent. When the reagent diffuses into the substrate it becomes bound to it, and a kind of cascade precipitation occurs. As shown in \cite{meinhardt}, a cascade of simple molecular interactions permits reliable pattern formation in an iterative way. 


Beyond their utilitarian use as fast-prototyping tools, simple systems such as CA that are capable of abstracting the substratum from the mechanism have the advantage of affording insights into the behaviour of systems independent of physical or chemical carriers. One may, for example, inquire into the importance of the right rates of systems if these reactions are to occur. As with seashells, statistical evidence and similarities between systems don't necessarily prove that the natural cases are cases of Turing pattern formation. They could actually be produced by some other chemical reaction, and just happen to look like Turing patterns. That pigmentation in animals is produced by Turing pattern formation is now generally accepted, but it is much more difficult to ascribe the origin of what we take to be patterns in certain other places to Turing-type formations. 

But as I will argue in the following section, a theoretical model based on algorithmic probability suggests that the simpler the process generating a pattern, the greater the chances of its being the kind of mechanistic process actually underlying Turing pattern formation. For there is a mathematical theory which asserts that among the possible processes leading to a given pattern, the simplest is the one most likely to be actually producing it. Processes similar to Turing pattern formation would thus be responsible for most of the patterns in the world around us, not just the special cases. Turing patterns would fit within this larger theory of pattern formation, which by the way is the acknowledged mathematical theory of patterns, even if it is not identified as such in the discourse of the discipline. On account of its properties it is sometimes called the miraculous distribution \cite{kirchherr}.

\section{Turing patterns with Turing machines}

As recently observed by \cite{brenner}, the most interesting connection to Turing's morphogenesis is perhaps to be found in Turing's most important paper on computation \cite{turing}, where the concept of computational universality is presented. We strengthen that connection in this paper by way of the concept of algorithmic probability.

 As has been pointed out by Cooper \cite{cooper}, Turing's model of computation \cite{turing} was the one that was closest to the mechanistic spirit of the Newtonian paradigm in science. The Turing machine offered a model of computability of functions over the range of natural numbers and provided a framework of algorithmic processes with an easy correspondence to physical structures. In other words, Turing's model brought with it a mechanical realisation of computation, unlike other approaches. 

Turing made another great contribution in showing that universality came at a cost, it being impossible to tell whether or not a given computation would halt without actually performing the computation (assuming the availability of as much time as would be needed). Turing established that for computational systems in general, one cannot say whether or not they would halt, thereby identifying a form of unpredictability of a fundamental nature even in fully deterministic systems. 

As Cooper points out \cite{cooper}, it is well-known in mathematics that complicated descriptions may take us beyond what is computable. In writing a program to tackle a problem, for example, one has first to find a precise way to describe the problem, and then devise an algorithm to carry out the computation that will yield a solution. To arrive at the solution then, one has to run a machine on the program that has been written. My colleagues and I have made attempts in the past to address epistemological questions using information theory and computer experiments, with some interesting results \cite{joost}. 

Computer simulations performed as part of research into artificial life have reproduced various known features of life processes and of biological evolution. Evolution seems to manifest some fundamental properties of computation, and not only does it seem to resemble an algorithmic process \cite{dennett,wolfram}, it often seems to produce the kinds of persistent structures and output distributions a computation could be expected to produce \cite{zenilalgo}. 

Among recent discoveries in molecular biology is the finding that genes form building blocks out of which living systems are constructed \cite{biobrick}, a discovery that sheds new light on the common principles underlying the development of organs that are functional components rather than mere biochemical ingredients. The theory of evolution serves as one grand organising principle, but as has been pointed out before \cite{mitchell,chaitin,chaitinbio,chaitinbio2}, it has long lacked a formal mathematical general theory, despite several efforts to supply this deficiency (e.g. \cite{hopfield,grafen,grafen2}). 

Recently there has been an interest in the ``shape'' of a self-assembled system as output of a computational process \cite{adleman,rothemund,adleman2,rothemund2,aggarwal}. These kinds of processes have been modelled using computation before \cite{winfree}, and the concept of self-assembly has been extensively studied by computer scientists since von Neumann \cite{neumann}, who himself studied features of computational systems capable of displaying persistent self-replicating structures as an essential aspect of life, notably using CA. Eventually these studies produced systems manifesting many features of life processes \cite{conway,langton,wolfram}, all of which have turned out to be profoundly connected to the concept of (Turing) universal computation (Conway's Game of Life, Langton's ant, Wolfram's Rule 110). Some artificial self-assembly models \cite{rothemund}, for example, demonstrate all the features necessary for Turing-universal computation and are capable of yielding arbitrary shapes \cite{winfree2} such as a Turing-universal biomolecular system. 

It is known that computing systems capable of Turing universality have properties that are not predictable in principle. In practice too it has been the case that years of various attempts have not yet yielded a formula for predicting the evolution of certain computing systems, despite their apparent simplicity. An example is the elementary CA Rule 30 \cite{wolfram}, by most measures the simplest possible computing rule exhibiting apparently random behaviour. 

So the question is how deeply pattern formation and Turing universality are connected. If reformulated, the answer to this question is not far to find. The Turing universality of a system simply refers to its capability of producing any possible pattern, and the question here as well is simply by what specific means (what programs), and how often these programs can be found to produce patterns in nature. The answer, again, can be derived from algorithmic information theory, particularly algorithmic probability, at the core of which is the concept of Turing universality.

Today we know that the necessary elements for (Turing) computational universality are minimal, as shown by Rule 110 and tag systems that have led to the construction of the smallest Turing machines capable of universal computation \cite{wolfram,cook,neary}. 

One way to connect Turing's theory of morphogenesis to his theory of computation is by way of the theory of algorithmic information, a theory that was developed at least a decade later \cite{kolmo,chaitin}, and wouldn't really be known (being rediscovered several times) until two decades later. Turing couldn't have anticipated such a connection, especially with regard to a possible generalisation of the problem he identified concerning the violation of parity after random symmetry breaking by (uniform) random disruptions, which would then lead to a uniform distribution of patterns---something that to us seems clearly not to be the case. As we will see, by introducing the concept of algorithmic probability we use a law of information theory that can account for important bias but not for symmetry violation. Hence while the theory may provide a solution to a generalised problem, it is likely that symmetry imbalance is due to a non-informational constraint, hence a physical, chemical or biological constraint that organisms have to reckon with in their biological development. The proposed generalisation reconnects Turing patterns to Turing machines.

 The algorithmic complexity $K_U(s)$ of a string $s$ with respect to a universal Turing machine $U$, measured in bits, is defined as the length in bits of the shortest (prefix-free\footnote{That is, a machine for which a valid program is never the beginning of any other program, so that one can define a convergent probability the sum of which is at most 1.}) Turing machine $U$ that produces the string $s$ and halts~\cite{kolmo,chaitin,levin,solomonoff}. Formally,

\begin{equation}
\label{kolmo}
\noindent K_U(s) = \min\{|p|, U(p)=s\}\\
\textit{ where $|p|$ is the length of $p$ measured in bits.}
\end{equation}

This complexity measure clearly seems to depend on $U$, and one may ask whether there exists a Turing machine which yields different values of $K_U(s)$ for different $U$. The ability of Turing machines to efficiently simulate each other implies a corresponding degree of robustness. The invariance theorem~\cite{solomonoff,chaitin} states that if $K_U(s)$ and $K_{U^\prime}(s)$ are the shortest programs generating $s$ using the universal Turing machines $U$ and $U^\prime$ respectively, their difference will be bounded by an additive constant independent of $s$. Formally:

\begin{equation}
\label{invariance}
|K_U(s) - K_{U^\prime}(s)| \leq c_{_{U,U^\prime}}
\end{equation}

Hence it makes sense to talk about $K(s)$ without the subindex $U$. $K(s)$ is lower semi-computable, meaning that it can be approximated from above, for example, via lossless compression algorithms. 

From equation \ref{kolmo} and based on the robustness provided by equation \ref{invariance}, one can formally call a string $s$ a Kolmogorov (or algorithmically) random string if $K(s) \sim |s|$ where $|s|$ is the length of the binary string $s$. Hence an object with high Kolmogorov complexity is an object with low algorithmic structure, because Kolmogorov complexity measures \emph{randomness}---the higher the Kolmogorov complexity, the more random. This is the sense in which we use the term \emph{algorithmic structure} throughout this paper---as opposed to randomness.

\subsection{Algorithmic probability}

As for accounting for unequal numbers of patterns, the notion of algorithmic probability, introduced by Ray Solomonoff \cite{solomonoff} and formalised by Leonid Levin \cite{levin}, describes the probability distribution of patterns when produced by a (computational) process. The algorithmic probability of a string $s$ is the probability of producing $s$ with a random program $p$ running on a universal (prefix-free) Turing machine. In terms of developmental biology, a prefix-free machine is a machine that cannot start building an organism, and having done so, begin building another one, because a valid (self-delimited) program describing the instructions for building a viable organism cannot contain another valid (self-delimited) program for building another one. Formally,

\begin{equation}
Pr(s) = \sum_{p : U(p) = s} 2^{-|p|}
\end{equation}

That is, the sum over all the programs for which the universal Turing machine $U$ with $p$ outputs the string $s$ and halts. $U$ is required to be a universal Turing machine to guarantee that the definition is well constructed for any $s$, that is, that there is at least a program $p$ running on $U$ that produces $s$. As $p$ is itself a binary string, $Pr(s)$ is the probability that the output of $U$ is $s$ when provided with a random program (with each of its program bits independent and uniformly distributed). 

$Pr(s)$ is related to algorithmic (Kolmogorov) complexity \cite{solomonoff,kolmo,levin,chaitin} in that the length of the shortest program (hence $K(s)$) is the maximum term in the summation of programs contributing to $Pr(s)$. But central to the concept of algorithmic emergence is the following (coding) theorem \cite{cover,calude}:

\begin{equation}
\label{eq}
K(s)=-log Pr(s)+O(1)\textit{ or simply }K(s) \sim -log Pr(s)
\end{equation}
An interpretation of this theorem is that if a string has many long descriptions it also has a short one \cite{downey}.

 In essence this coding theorem asserts that the probability of a computer producing the string $s$ when fed with a random program running on a universal Turing machine $U$ is largely determined by $K(s)$. This means that outputs having low algorithmic complexity (`structured' outputs) are highly probable, while random looking outputs (having high algorithmic complexity) are exponentially less likely. The fact that $Pr(s) \approx 2^{-K(s)}$ is non-obvious because there are an infinite number of programs generating $s$, so a priori $s$ may have had high universal probability due to having been generated by numerous long programs. But the coding theorem \cite{calude,cover} shows that if an object has many long descriptions, it must also have a short one. 

To illustrate the concept, let's say that a computation keeps producing an output of alternating 1s and 0s. Algorithmic probability would indicate that, if no other information about the system were available, the best possible bet is that after $n$ repetitions of $01$ or $10$ the computation will continue in the same fashion. Therefore it will produce another $01$ or $10$ segment. In other words, patterns are favoured, and this is how the concept is related to algorithmic complexity---because a computation with low algorithmic randomness will present more patterns. And according to algorithmic probability, a machine will more likely produce and keep (re)producing the same pattern.

 Algorithmic probability as a theory of patterns has the advantage of assuming very little, and stands out as a particularly simple way to illustrate the general principle of pattern formation.

\subsection{Homogeneity and the breakdown of symmetry}

The transition rules described in Turing's paper with the ODEs would allow the cells to ``communicate'' via diffusion of the chemicals, and the question in the context of symmetry was whether the aggregation of cells communicating via diffusion would remain in an homogeneous resting state. The model indicates that depending upon the chemical reactions and the nature of the diffusion, the aggregation of cells (e.g. a tissue) would be unstable and would develop patterns as a result of a break in the symmetry of chemical concentrations from cell to cell. 

The development of multicellular organisms begins with a single fertilised egg, but the unfolding process involves the specification of diverse cells of different types. Such an unfolding is apparently driven by an asymmetric cell division crucial in determining the role of each cell in the organism's body plan. Pattern formation occurs outside of equilibrium at the centre of this kind of symmetry breaking. 

In the real world, highly organised strings have little chance of making it if they interact with other systems, because symmetry is very weak. Yet not only do structures persist in the world (otherwise we would only experience randomness), but they may in fact be generated by symmetry breaking. Changing a single bit, for example, destroys a perfect 2-period pattern of a $(01)^n$ string, and the longer the string, the greater the odds of it being destroyed by an interaction with another system. But a random string will likely remain random-looking after changing a single bit.

 One can then rather straightforwardly derive a thermodynamical principle based on the chances of a structure being created or destroyed. By measuring the Hamming distance between strings of the same length, we determine the number of changes that a string needs to undergo in order to remain within the same complexity class (the class of strings with identical Komogorov complexity), and thereby determine the chances of its conserving structure versus giving way to randomness. If $H$ is the function retrieving the Hamming distance, and $s=$010101 is the string subject to bit changes, it can only remain symmetrical under the identity or after a $H(010101,101010)= 6$ bit-by-bit transformation to remain in the same low Kolmogorov (non-random) complexity class (the strings are in the same complexity class simply because one is the reversion of the other). On the other hand, a more random-looking string of the same length, such as 000100, only requires $H(000100,001000)=2$ changes to remain in the same high (random-looking) Kolmogorov complexity class. In other words, the shortest path for transforming $s=$010101 into $s^\prime=$101010 requires six changes to preserve its complexity, while the string 000100 requires two changes to become 001000 in the same complexity class. It is clear that the classical probability of six precise bit changes occurring is lower than the probability of two such changes occurring. Moreover, the only chance the first string 010101 has of remaining in the same complexity class is by becoming the specific string 101010, while for the second string 000100, there are other possibilities: 001000, 110111 and 111011. In fact it seems easier to provide an informational basis for thermodynamics than to explain how structures persist in such an apparently weak state of the world. 

But things are very different where the algorithmic approach of low-level generation of structure is concerned. If it is not a matter of a bit-by-bit transformation of strings but rather of a computer program that generates $s^\prime$ out of $s$ (its conditional Kolmogorov complexity), that is $K(s^\prime,s)$ then the probabilities are very different, and producing structured strings turns out to actually be more likely than producing random-looking ones. So, for example, if instead of changing a string bit by bit one uses a program to reverse it, the reversing program is of fixed (short) length and will remain the same size in comparison to the length of the string being reversed. This illustrates the way in which algorithmic probability differs from classical probability when it comes to symmetry preserving and symmetry breaking.

 Algorithmic probability does not solve Turing's original problem, because a string and its reversed or (if in binary) complementary version have the same algorithmic probability (for a numerical calculation showing this phenomenon and empirically validating Levin's coding theorem, see \cite{delahayezenil}). The weakness of symmetrical strings in the real world suggests that symmetric bias is likely to be of physical, chemical or even biological origin, just as Turing believed.

\subsection{A general algorithmic model of structure formation}

When we see patterns/order in an object (e.g. biological structures), we tend to think they cannot be the result of randomness. While this is certainly true if the parts of the object are independently randomly generated, the coding theorem shows that it is not true if the object is the result of a process (i.e. computation) which is fed randomness. Rather if the object is the result of such a process, we should fully expect to see order and patterns. 

Based on algorithmic probability, Levin's universal distribution \cite{levin} (also called Levin's semi-measure) describes the expected pattern frequency distribution of an abstract machine running a random program relevant to my proposed research program. A process that produces a string $s$ with a program $p$ when executed on a universal Turing machine $U$ has probability $Pr(s)$. 

For example, if you wished to produce the digits of the mathematical constant $\pi$ by throwing digits at random, you'd have to try again and again until you got a few consecutive numbers matching an initial segment of the expansion of $\pi$. The probability of succeeding would be very small: $1/10$ to the power of the desired number of digits in base 10. For example, $(1/10)^{5000}$ for a segment of only length 5000 digits of $\pi$. But if instead of throwing digits into the air one were to throw bits of computer programs and execute them on a digital computer, things turn out to be very different. A program producing the digits of the expansion of $\pi$ would have a greater chance of being produced by a computer program shorter than the length of the segment of the $\pi$ expected. The question is whether there are programs shorter than such a segment. For $\pi$ we know they exist, because it is a (Turing) computable number. 

Consider an unknown operation generating a binary string of length $k$ bits. If the method is uniformly random, the probability of finding a particular string $s$ is exactly $2^{-k}$, the same as for any other string of length $k$, which is equivalent to the chances of picking the right digits of $\pi$. However, data (just like $\pi$---largely present in nature, for example, in the form of common processes relating to curves) are usually produced not at random but by a specific process. 

But there is no program capable of finding the shortest programs (due to the non-computability of $Pr(s)$ \cite{levin,chaitin}), and this limits the applicability of such a theory. Under this view, computer programs can in some sense be regarded as physical laws: they produce structure by filtering out a portion of what one feeds them. And such a view is capable of explaining a much larger range, indeed a whole world of pattern production processes. Start with a random-looking string and run a randomly chosen program on it, and there's a good chance your random-looking string will be turned into a regular, often non-trivial, and highly organised one. In contrast, if you were to throw particles, the chances that they'd group in the way they do if there were no physical laws organising them would be so small that nothing would happen in the universe. It would look random, for physical laws, like computer programs, make things happen in an orderly fashion.

Roughly speaking, $Pr(s)$ establishes that if there are many long descriptions of a certain string, then there is also a short description (low algorithmic complexity), and vice versa. As neither $C(s)$ nor $Pr(s)$ is computable, no program can exist which takes a string $s$ as input and produces $Pr(s)$ as output. Although this model is very simple, the outcome can be linked to biological robustness, for example to the distribution of biological forms (see \cite{zenilalgo,zenilmarshall}). We can predict, for example, that random mutations should on average produce (algorithmically) simpler phenotypes, potentially leading to a bias toward simpler phenotypes in nature \cite{kamalzenil}.

The distribution $Pr(s)$ has an interesting particularity: it is the process that determines its shape and not the initial condition the programs start out from. This is important because one does not make any assumptions about the distribution of initial conditions, while one does make assumptions about the distribution of programs. Programs running on a universal Turing machine should be uniform, which does not necessarily mean truly random. For example, to approach $Pr(s)$ from below, one can actually define a set of programs of a certain size, and define any enumeration to systematically run the programs one by one and produce the same kind of distribution. This is an indication of robustness that I take to be a salient property of emergence, just as it is a property of many natural systems.

 Just as strings can be produced by programs, we may ask after the probability of a certain outcome from a certain natural phenomenon, if the phenomenon, just like a computing machine, is a process rather than a random event. If no other information about the phenomenon is assumed, one can see whether $Pr(s)$ says anything about a distribution of possible outcomes in the real world. In a world of computable processes, $Pr(s)$ would give the probability that a natural phenomenon produces a particular outcome and indicate how often a certain pattern would occur. If you were going to bet on certain events in the real world, without having any other information, $Pr(s)$ would be your best option if the generating process were algorithmic in nature and no other information were available.

\subsection{Algorithmic complexity and one-dimensional patterns}

It is difficult to measure the algorithmic (Kolmogorov) complexity of a Turing pattern if given by an ODE. However, because we know that these patterns can be simulated by CAs, one can measure their algorithmic complexity by the shortest CA program producing that pattern running on a universal Turing machine. CAs turn out to have low algorithmic complexity because programs capable of such complexity can be described in a few bits. More formally, we can say that the pattern is not random because it can be compressed by the CA program generating it. While the pattern can grow indefinitely, the CA program length is fixed. On the other hand, use of the coding theorem also suggests that the algorithmic complexity of a Turing pattern is low, because there is the empirical finding that many programs are capable of producing similar, if not the same low complexity patterns. For example, among the 256 ECA, about two thirds produce trivial behaviour resulting in the same kind of low complexity patterns. The bias towards simplicity is clear. This means that it is also true that the most frequent patterns are those which seem to have the lowest algorithmic complexity, in accordance with the distribution predicted by the coding theorem. For further information these results have been numerically quantified using a compressibility index in \cite{zenilcompress} approaching the algorithmic (Kolmogorov) complexity of ECAs.

It is possible to investigate one-dimensional patterns produced by Turing machines. We have run all (more than $11\times10^9$) Turing machines with 2 symbols and up to 4 states \cite{delahayezenil} thanks to the fact that we know the halting time of the so-called busy beaver Turing machines. After counting the number of occurrences of strings of length 9 produced by all these TMs (see the top 10 occurrences in Table 1) we find that, as the coding theorem establishes, the most frequent strings have lower random (Kolmogorov) complexity, what one would easily recognise as a ``one-dimensional'' Turing sort of pattern. The ones at the top are patterns that we do not find at all interesting, and are among those having the lowest Kolmogorov complexity. But among skin types in nature, for example,  monochromatic skins, such as in many species of ducks or swans, elephants, dolphins and Hominidae (including humans) are common, even if we pay more attention to skin patterns such as stripes and spots. Among the top 10 of the $2^9$ possible strings, one can see strings that correspond to recognisable patterns, evidently constrained by their one-dimensional nature, but along the lines of what one would expect from ``one-dimensional'' patterns as opposed  to what one would see in ``two-dimensional'' (e.g. skin). A one-dimensional spotted or striped string is one of the form ``01" repeated $n$ types.

\begin{table}[htdp]
\caption{The 10 most frequent one-dimensional patterns produced by all 4-state 2-symbol Turing machines starting from empty input (for full results see \cite{delahayezenil}), hence an approximation of $Pr(s)$, the algorithmic probability that we have denoted by $Pr_4(s)$, that is the outcome frequency/multiplicity of the pattern. The visual representation shows how the pattern would look like if used as a skin (Turing-type) motif.}
\begin{center}
\begin{tabular}{|r|c|r|}
\hline
output & visual & frequency \textit{ }  \\
string ($s$) & representation & $Pr_4(s)$ \textit{ }  \\
\hline
000000000& \scalebox{.26}{\includegraphics{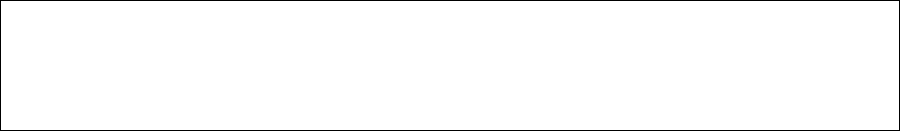}} & $1.3466\times10^{-7}$\\
111111111&\scalebox{.26}{\includegraphics{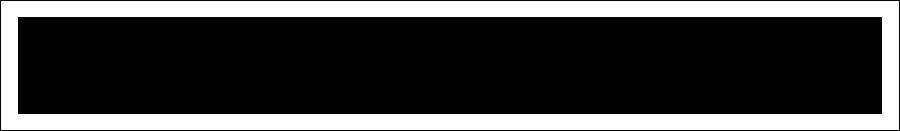}} & $1.3466\times10^{-7}$\\
000010000& \scalebox{.26}{\includegraphics{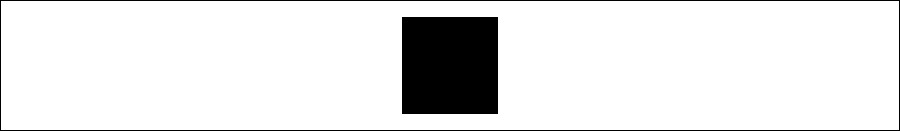}}& $7.83899\times10^{-8}$\\
111101111& \scalebox{.26}{\includegraphics{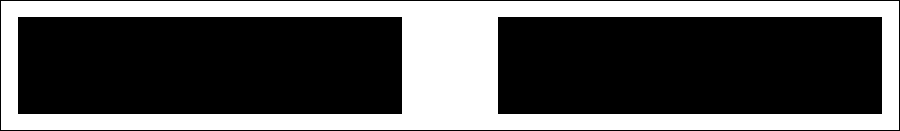}}& $7.83899\times10^{-8}$\\
000000001&\scalebox{.26}{\includegraphics{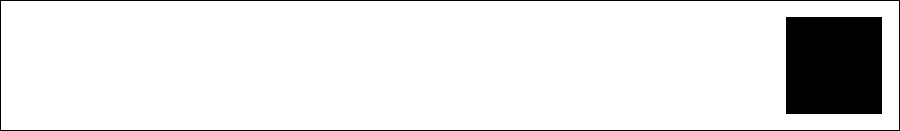}} & $7.53699\times10^{-8}$\\
011111111& \scalebox{.26}{\includegraphics{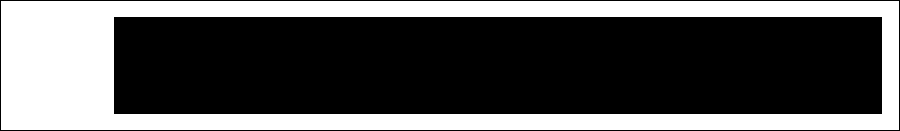}}& $7.53699\times10^{-8}$\\
100000000& \scalebox{.26}{\includegraphics{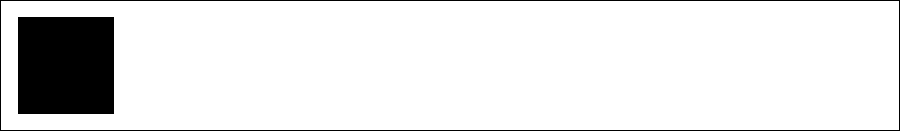}}& $7.53699\times10^{-8}$\\
111111110& \scalebox{.26}{\includegraphics{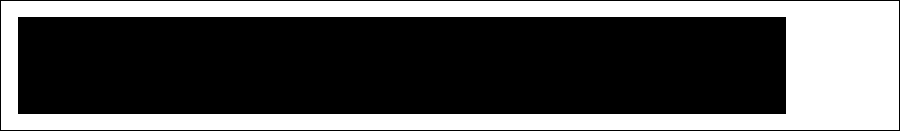}}& $7.53699\times10^{-8}$\\
010101010& \scalebox{.26}{\includegraphics{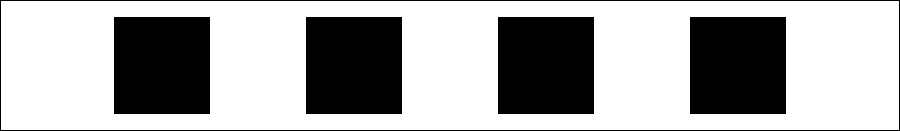}}& $4.422\times10^{-8}$\\
101010101& \scalebox{.26}{\includegraphics{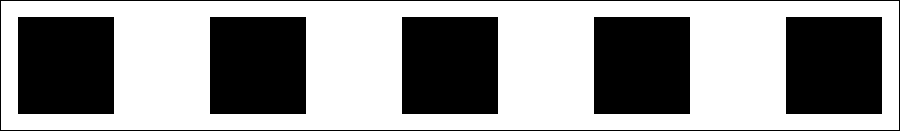}}& $4.422\times10^{-8}$\\
\hline
\end{tabular}
\end{center}
\label{default}
\end{table}%

To produce Turing-type patterns with Turing machines in order to study their algorithmic complexity would require two-dimensional Turing machines, which makes the calculation very difficult in practice due to the combinatorial explosion of the number of cases to consider. This difficulty can be partially overcome by sampling, and this is a research project we would like to see developed in the future, but for the time being it is clear that the one-dimensional case already produces the kinds of patterns one would expect from a low level general theory of pattern emergence, encompassing Turing-type patterns and rooted in Turing's computational universality.

\section{Concluding remarks}

It seems that one point is now very well established, viz. that universality can be based on extremely simple mechanisms. The success of applying ideas from computing theories to biology should encourage researchers to go further. Dismissing this phenomenon of ubiquitous universality in biology as a mere technicality having little meaning is a mistake. 


In this paper, we have shown that what algorithmic probability conveys is that structures like Turing patterns will likely be produced as the result of random computer programs, because structured patterns have low algorithmic complexity (are more easily described). A simulation of two-dimensional Turing machines would produce the kind of patterns that Turing described, but because running all Turing machines to show how likely this would be among the distribution of possible patterns is computationally very expensive, one can observe how patterns emerge by running one-dimensional Turing machines.

 As shown in \cite{delahayezenil}, in effect the most common patterns are the ones that look less random and more structured. In \cite{zenilalgo} it is also shown that different computational models produce reasonable and similar frequency distributions of patterns, with the most frequent outputs being the ones with greater apparent structure, i.e. with lower algorithmic (Kolmogorov) complexity \cite{kolmo}. It is important to point out, however, that we are not suggesting that this low level process of structure generation substitutes for any mechanism of natural selection. We are suggesting that computation produces such patterns and natural selection picks those that are likely to provide an evolutionary advantage. If the proportion of patterns (e.g. skin patterns among living systems) are distributed as described by $Pr(s)$, it may be possible to support this basic model with statistical evidence.

 We think this is a reasonable and promising approach for arriving at an interesting low level formalisation of emergent phenomena, using a purely computational framework. It is all the more interesting in that it is capable of integrating two of Alan Turing's most important contributions to science: his concept of universal computation and his mechanism of pattern formation, grounding the latter in the former through the use of algorithmic probability.



\end{document}